\documentclass[12pt]{article}
\usepackage{amsmath,amssymb,amsthm,color}
\usepackage{graphicx}
\usepackage{cite}
\usepackage{epsfig}
\renewcommand{\baselinestretch}{2.4}
\begin{document}
%\begin{frontmatter}
%
\title{Substrate-induced structures of bismuth adsorption on graphene: a first principle study \\}
\author{Shih-Yang Lin,$^a$ Shen-Lin Chang,$^b$ Hsin-Hsien Chen,$^a$ \\ Shu-Hsuan Su,$^a$ Jung-Chun Huang,$^{*,a,c,d}$ Ming-Fa Lin$^{*,a}$
\small  $$\\
\small  $^a$Department of Physics, National Cheng Kung University, Tainan 701, Taiwan. \\
\small  $^b$Department of Electrophysics, National Chiao Tung University, Hsinchu 300, Taiwan. \\
\small  $^c$Advanced Optoelectronic Technology Center, National Cheng Kung University, Tainan 701, Taiwan. \\
\small  $^d$Taiwan Consortium of Emergent Crystalline Materials, Ministry of Science and Technology, \\
\small  Taipei 106, Taiwan. \\
 }
\renewcommand{\baselinestretch}{1}
\maketitle

\renewcommand{\baselinestretch}{1.4}
\begin{abstract}
The geometric and electronic properties of Bi-adsorbed monolayer graphene, enriched by the strong effect of substrate, are investigated by first-principles calculations. The six-layered substrate, corrugated buffer layer, and slightly deformed monolayer graphene are all simulated. Adatom arrangements are thoroughly studied by analyzing the ground-state energies, bismuth adsorption energies, and Bi-Bi interaction energies of different adatom heights, inter-adatom distance, adsorption sites, and hexagonal positions. A hexagonal array of Bi atoms is dominated by the interactions between the buffer layer and the monolayer graphene. An increase in temperature can overcome a $\sim 50$ meV energy barrier and induce triangular and rectangular nanoclusters. The most stable and metastable structures agree with the scanning tunneling microscopy measurements. The density of states exhibits a finite value at the Fermi level, a dip at $\sim -0.2$ eV, and a peak at $\sim -0.6$ eV, as observed in the experimental measurements of the tunneling conductance.
\vskip 1.0 truecm
\par\noindent  * Corresponding author. ~{{\it E-mail addresses}:\\ jcahuang@mail.ncku.edu.tw) (J. C. A. Huang), mflin@mail.ncku.edu.tw (M.F. Lin)}
\end{abstract}

\pagebreak
\renewcommand{\baselinestretch}{2}
\newpage

{\bf 1. Introduction}
\vskip 0.3 truecm

Two-dimensional graphene has been a main-stream material in both fundamental research and device applications since 2004,\cite{novoselov2004electric,novoselov2005two,geim2007rise,he2012graphene}
mainly owing to its remarkable physical, chemical and material properties.\cite{zhang2005experimental,novoselov2006unconventional,orlita2008approaching,wang2009high,balandin2008superior,lee2008measurement,vadukumpully2011flexible,min2008room,perali2013high,barbarino2015intrinsic} Graphene possesses a unique carrier mobility,\cite{orlita2008approaching,wang2009high} thermal conductivity,\cite{balandin2008superior,barbarino2015intrinsic} mechanical strength,\cite{lee2008measurement,vadukumpully2011flexible} a quantum Hall effect,\cite{zhang2005experimental,novoselov2006unconventional} and superfluidity.\cite{min2008room,perali2013high} Monolayer graphene is a zero-gap semiconductor since the Dirac points have a vanishing density of states (DOS) at the Fermi level ($E_{F}$). The electronic properties can be tuned by changing the doping elements,\cite{ryu2010atmospheric,elias2009control,yeh2014nitrogen,krasheninnikov2009embedding,marchenko2012giant,schultz2013chemical} geometric curvatures,\cite{guinea2008midgap,lin2015feature} layer numbers,\cite{sutter2009electronic,hao2010probing} stacking configurations,\cite{hao2010probing,lin2015magneto} mechanical strain,\cite{guinea2010energy,wong2012strain} and the applied electric and magnetic fields.\cite{castro2007biased,lai2008magnetoelectronic} Adatom-adsorbed graphenes can be successfully produced in laboratory, such as wide-gap hydrogenated graphenes\cite{elias2009control} and metallic lithium-doped graphenes with a high free carrier density.\cite{mandeltort2012rapid} They are critically important in applications of nanoelectronic, nanophotonic, and energy storage devices. This work shows how the energetically favorable distribution configuration of Bi-adatoms can be determined by the critical interactions among the substrate, buffer layer, monolayer graphene, and adatoms.

The 2p$_{z}$ orbitals of carbon atoms are perpendicular to the graphene surface, providing a potential chemical environment for doping with various elements with the strong carbon-adatom bondings. Experimental research studies have successfully modulated the essential properties of graphene by using dopants, such as O,\cite{ito2008semiconducting,ryu2010atmospheric} H,\cite{elias2009control,ryu2008reversible} N,\cite{qu2010nitrogen,yeh2014nitrogen} transition metals,\cite{krasheninnikov2009embedding,pi2009electronic} and heavy metals.\cite{machida2006lead,marchenko2012giant} These doping elements induce the novel physical properties, including configuration transformations, semiconductor-metal transition, band-gap tuning, adatom-related DOS, creation of magnetic moments, and modulated spintronics. On the other hand, previous theoretical studies on adatom adsorptions revealed that the electronic properties are altered dramatically under various concentrations and distributions.\cite{chan2008first,khomyakov2009first,balog2010bandgap} Moreover, uniform or aggregated adatom distributions play an important role in tuning the electronic properties even at the same concentration.\cite{balog2010bandgap} Apparently, the special distribution of adatoms becomes a critical issue in fabricating high-quality nanoelectronic devices.

To date, bismuth-related systems are some of the most widely studied materials, since the dimensionality can enrich the fundamental properties. Bulk bismuth with rhombohedral symmetry is a semimetal with a long Fermi wavelength and small effective electron mass.\cite{hofmann2006surfaces} Its surface states belong to a Dirac fermion gas.\cite{li2008phase} The three bismuth surfaces: Bi (111), Bi (100), and Bi (110) have a higher free carrier density at $E_{F}=0$ compared to those of the bulk system.\cite{hofmann2006surfaces} The Bi thin film, in which the surface does not have strong chemical reactions with O$_{2}$, appears to be stable up to about $600$ K.\cite{bobaru2012competing} Furthermore, a 1D Bismuth nanowire displays narrow band gaps due to significant quantum confinement effects.\cite{black2002infrared} Moreover, bismuth has been explored in the fields of environmental engineering, biochemistry, and energy engineering. For example, Bi-based nanoelectrode arrays are used to detect heavy metals,\cite{wanekaya2011applications} a polycrystalline bismuth oxide film can serve as biosensor\cite{shan2009polycrystalline}, and bismuth oxide on nickel foam covered with thin carbon layers is used as the anode in lithium battery.\cite{li2013bismuth}

Recently, bismuth adatoms on graphene formed on a 4H-SiC (0001) substrate were clearly observed at room temperature.\cite{chen2015long,chen2015tailoring} The corrugated substrate and monolayer graphene layers have been identified by scanning tunneling microscopy (STM). Specifically, a large-scale hexagonal array of Bi atoms was revealed at room temperature. Such adatoms are aggregated in triangular and rectangular nanoclusters of uniform size under the further annealing process. The scan tunneling spectroscopy (STS) measurements of the dI/dV spectrum confirm the Dirac point, free conduction electrons, and Bi-related structures. These results shed light on controlling the nucleation of adatoms and subsequent growth of nanostructures on graphene surfaces. On the other hand, there are some theoretical studies on the geometric structures and energy bands of Bi-adsorbed and Bi-intercalated graphenes.\cite{hsu2013first,akturk2010bismuth} The former have been performed on monolayer graphene without simulation of the substrate and the buffer graphene layer, and thus the deformed graphene surface structure may be unreliable.\cite{akturk2010bismuth} The latter is calculated for Bi and/or Sb as a buffer layer above the four-layer SiC substrate, which results in an energetically unfavorable enviroment for the metal atoms to be adsorbed on the graphene sheet.\cite{hsu2013first} A comprehensive study on the critical roles played by the configuration of the Bi adatoms, buffer layer, and substrate is absent to date.

How the energetically favorable distribution configuration of Bi adatoms can be determined by the substrate effect and the buffer layer is explored in detail by the first-principles calculations. This work shows that the adsorption energy strongly depends on the various atomic sites and hexagonal positions. The former include hollow, top, and bridge sites, and the latter are closely related to the non-uniform Van der Walls interactions. Two kinds of different bismuth distributions could be obtained, namely uniform or aggregated adatom arrangements with critical energy barriers induced by the buffer layer. The aggregated adatom arrangements are associated with the large-scale hexagonal symmetry, Bi coverage, and relatively few vacancies. The optimized geometric structures were validated by the STM measurements.\cite{chen2015long,chen2015tailoring} Hopefully, these rich fundamental features in bismuth structures can promote potential applications for use in electronic devices and energy materials. Moreover, the main effects of the Bi-adsorption on the density of states (DOS) could be understood through a exhaustive comparison with tunneling conductance measurements.\cite{chen2015long,chen2015tailoring}

%\end{document}
\vskip 0.6 truecm
\par\noindent
{\bf 2. Methods }
\vskip 0.3 truecm

The first-principles calculations are based on the density functional theory (DFT) implemented by the Vienna \emph{ab initio} simulation package.\cite{kresse1996efficient} The generalized gradient approximation, within the Perdew-Burke-Ernzerhof functional,\cite{perdew1996generalized} is applied to describe the exchange-correlation energy of interacting electrons. The projector augmented wave is utilized to characterize the electron-ion interactions. The Van der Waals force is included in the calculations using the semiemprical DFT-D2 correction of Grimme to correctly represent the atomic interactions between graphene layers.\cite{grimme2006semiempirical} A vacuum space of 15 {\AA} is inserted between periodic images to avoid their interactions. The cutoff energies of the wave function expanded by plane waves are chosen to be 400 eV. For the calculations of the electronic properties and the optimal geometric structures, the first Brillouin zones are sampled by  $3\times3\times1$ \emph{k}-points via the Monkhorst-Pack scheme. The convergence of the Helmann-Feymann force is set to be 0.01 eV ${\AA}^{-1}$.

\vskip 0.6 truecm
\par\noindent
{\bf 3. Results and discussion}
\vskip 0.3 truecm

First, in order to accurately simulate the bismuth-adsorbed monolayer graphene, the six-layer Si-terminated 4H-SiC (0001) substrate is taken into account. The optimized results show that the four-layer substrate (region I in Fig. 1), which is reduced from the six-layer one, has almost identical geometric properties to the six-layered one. Second, a buffer layer, as shown in region II of Fig. 1, is in a periodic ripple shape after relaxation, in which the troughs bond with the silicon atoms of the substrate. This periodic corrugation is almost identical to that revealed by the STM measurements.\cite{chen2015long} Third, the monolayer graphene is nearly flat with a slightly extended C-C bond length of 1.50 {\AA}, as indicated in region III. The interlayer distance between the monolayer graphene and the buffer layer varies from 3.21 {\AA} to 5.45 {\AA} at the crests and troughs, respectively. This difference clearly illustrates the non-uniform Van der Waals interactions between them, and thus dominates the distribution of the Bi adatoms. Finally, bismuth atoms can be adsorbed on monolayer graphene in self-consistent calculations (region IV in Fig. 1). Up to now, based on the different experimental environments, two kinds of adatom distributions have been observed, namely a uniform hexagonal distribution and bismuth nanoclusters; these two distributions are depended on the adsorption energies.

The adsorption energy $\Delta E$, characterizing the reduced energy due to the bismuth-adsorption on graphene, is very useful for understanding the optimized geometric structure. It is defined as
\begin{eqnarray}
\Delta E = E_{Total}-E_{Gra}-E_{Buf}-E_{Bi} ,
\end{eqnarray}
where $E_{Total}$, $E_{Gra}$, $E_{Buf}$, and $E_{Bi}$ are the total energies of the composite system, pristine monolayer graphene, buffer layer, and isolated bismuth atoms, respectively. Three adsorption sites with higher geometric symmetry are investigated, i.e., the hollow, bridge, and top sites (Table 1). The bismuth atom on the hollow site of the carbon hexagon has the smallest adsorption energy, meaning that this configuration is less stable. The bridge and top sites have comparable adsorption energies with the former possessing the largest adsorption energy. This suggests that bismuth atoms are most likely to be observed at bridge sites. Moreover, the optimal distance $h$ between the adatoms and graphene surface is distinct for the different adsorption sites. The shorter distance at the bridge sites indicates stronger interactions between them, and is responsible for the structural stability. The adatom height of 2.32 {\AA} is consistent with that of STM measurements.\cite{chen2015long}

\begin{table}[h]
\small
\caption{Adsorption energies and heights for various atomic sites.}
\label{t1}
  \begin{tabular*}{0.5\textwidth}{@{\extracolsep{\fill}}lll}
\hline
site & $\Delta$E (eV) & h (\AA)\\
\hline
hollow & 0.6602 & 2.51 \\
bridge & 1.3450 & 2.32 \\
top & 1.2904 & 2.34 \\
\hline
\end{tabular*}
\end{table}

The distribution of bismuth atoms can be clearly explored by calculating the ground-state energy. The bridge site is a much more stable one within a single hexagonal ring, according to the above-mentioned results. The ground-state energies for the bridge sites in different hexagons along the periodic armchair direction are all calculated, these energies provide the full information needed to determine the most stable position in monolayer graphene, as shown in Fig. 2. These positions can be further divided into the three regions, red, yellow, and gray ones. The red hexagonal region is closest to the buffer layer with a distance of 3.21 {\AA}; it corresponds to the lowest ground-state energy among all bridge sites. This energy is set to zero in order to compare it with the other bridge sites. A bit away from the red region, the bridge sites between red and yellow sticks have higher ground-state energies. The energy differences is about $17-23$ meV. The other parts associated with the gray hexagons, which possesses ground-state energies comparable to each other, exhibit the highest energy difference in the range of $48$-$52$ meV. The energy variations clearly illustrate that the bismuth atoms are hardly transported from the red to the other regions at room temperature. The Bi atoms are most stable at the red hexagonal rings, since at these sites the Van der Walls interactions with the crests of the buffer layer are strongest. The energy barrier of $\sim50$ meV, which creates a potential well for the transportation of Bi adatoms, plays an important role in the dramatic change of adatom distribution during the variation of temperature.

The adsorption and ground-state energies are useful to comprehend the large-scale pattern of Bi adatoms identified by experimental measurements. The periodic hexagon image of Bi-adsorbed monolayer graphene is shown in Fig. 3(a), where the red and grey regions represent the most stable and unstable positions, respectively. The shortest, longest, and average distances between two Bi adatoms are, respectively, $14.2$ {\AA}, $18.1$ {\AA}, and $15.9$ {\AA}. The latter two are indicated by brown and black arrows between each red regions. This means that the most stable interatomic distance lies in the range of $14.2$-$18.1$ {\AA}. The simulated pattern deserves closer examination with the STM measurements,\cite{chen2015long,chen2015tailoring} which reveals the distribution of the Bi adatoms. The large-scale hexagonal array of Bi atoms is clearly identified in Fig. 3(b). Via the statistical analysis, it is determined that the interatomic distance is $15$ {\AA} and $16$ {\AA} for most of Bi adatoms, but $14$ {\AA} and $17$ {\AA} for some others (inset in Fig. 3(b)). All the simulation results agree with these STM measurements. This further illustrates that the periodic corrugation of the buffer layer indeed influences the monolayer graphene, and thus the arrangement of the Bi adatoms. The stable uniform distribution in room temperature should be very useful in the future applications in energy engineering.

Besides the most stable structure, there are metastable nano-structures induced by the annealing treatment.\cite{chen2015tailoring} As the temperature of Bi-adsorbed graphene is raised from $300$ K to $500$ K and then reduces to the original one, During the treatment, many Bi-adatom nanoclusters are observed in Fig. 4(a). Most of these are distributed in triangular and rectangular arrangements. These special configurations can be understood by a look at the ground-state energies and Bi-Bi interaction energies of optimized structures, as shown in Table 2. The Bi adatoms are set at the bridge sites of a hexagonal ring to get stronger interactions, as indicated in Fig. 4(b) for various adatom numbers (from 1 to 6). The higher the number of adatoms is, the lower $E_{total}$. The stronger attractive Bi-Bi interactions, compared to the Bi-C bonds, are responsible for this result. The latter, $E_{Bi-C}=-1.01$ eV, is obtained by subtracting the $E_{Total}$ of graphene from that of single-Bi-adsorbed graphene. The reduced energy due to the Bi-Bi interactions can be calculated by
\begin{eqnarray}
\Delta E_{Bi-Bi} = (E_{Total}-E_{Gra}-E_{Buf}-nE_{Bi}- \sum\limits_{i=1}^{n} E_{Bi-C})/n .
\end{eqnarray}
Among the various Bi-adatom numbers, $\Delta E_{Bi-Bi}$'s in the 3-, 4-, 5-, and 6-adatom nanoclusters are much lower than that of the 2-adatom one. This indicates that the former four are metastable structures. The experimental measurements show that most of the nanoclusters are composed of 3 and 4 adatoms (green and yellow arrows in Fig. 4(a)). If the bismuth coverage is insufficient, the 5- and 6-adatom nanoclusters are absent. A critical factor is that the bismuth atoms from the large-scale hexagonal array need to overcome the energy barrier ($\sim50$ meV in Fig. 2) in order to form these patterns. The temperature increase causes the transport of Bi adatoms between two neighboring unit cells (red hexagonal in Fig. 3(a)), where they have the same probability of moving toward or away from the hexagonal unit cell. The hexagonal symmetry leads to the 4-adatom-dominated nanoclusters. However, the relatively few vacancies in the large-scale hexagonal array (arrows in Fig. 3(b)) can create a the non-uniform transport environment and thus the 3-adatom nanoclusters. It should also be noted that the nearest distance between two bismuth clusters is about $16$ {\AA}, revealing that the energetic favorable adsorption sites of them correspond to the red hexagonal rings in Fig. 2. This further illustrates that the buffer layer plays an important role at various temperatures.

\begin{table}
\small
\caption{Total energies and Bi-Bi interaction energies for various Bi-nanoclusters.}
\label{t1}
  \begin{tabular*}{0.5\textwidth}{@{\extracolsep{\fill}}lll}
\hline
$\#$ Bi atoms & $E_{Total}$ (eV)& $\Delta E_{Bi-Bi}$ (eV) \\
\hline
1 & -2317.86 & x \\
2 & -2322.14 & -1.13 \\
3 & -2326.77 & -1.62 \\
4 & -2330.60 & -1.67 \\
5 & -2334.19 & -1.65 \\
6 & -2337.86 & -1.65 \\
\hline
\end{tabular*}
\end{table}

The density of states, shown in Figs. 5(a) and 5(b), can directly reflect the primary electronic properties. For a hexagonal array of Bi-adsorbed graphene, DOS is finite at $E=0$; it possesses a dip at low energy, and a peak structure at $E \sim -0.6$ eV (Fig. 5(a)). The first feature means that there exists a certain amount of free carriers. The second one at $-0.2$ eV is ascribed to the Dirac point. The third one comes from the contribution of bismuth atoms. Such electronic states are critical characteristics to identify the existence of bismuth atoms. The STS  measurements, in which the tunneling differential conductance map of the dI/dV-V curve is proportional to the DOS, can provide an accurate and efficient way to examine theoretical calculation. Previous experiments\cite{chen2015long,chen2015tailoring} show features similar to the DFT calculation, as shown in Fig. 5(b). The finite DOS at $E=0$ and the small dip at low energy show good agreement with the theoretical results. The peak at $\sim -0.7$ V originating from the bismuth atoms is similar to the peak structure in our calculation. The above-mentioned comparison of electronic properties might promote potential application in electronic devices.

\vskip 0.6 truecm
\par\noindent
{\bf 4. Conclusion }
\vskip 0.3 truecm

The geometric and electronic properties of bismuth adsorbed on monolayer graphene are studied by \emph{ab initio} density functional theory calculations. All the critical interactions are calculated, namely the six-layered substrate, periodically corrugated buffer layer, slightly deformed monolayer graphene, and adatom arrangements. The ground-state energies, bismuth adsorption energies, and Bi-Bi interaction energies for various Bi adatom configurations are investigated in detail. Remarkably, the corrugated buffer layer has periodic non-uniform Van der Walls interactions with the monolayer graphene. Adatom distributions are shown to be enriched by the substrate and the rippled buffer layer. The optimized structure, with an inter-adatom distance of $14 \sim 17$ {\AA}, presents a large-scale hexagonal array. The energy barrier for the formation of the adatom nanoclusters is about $50$ meV. With increasing temperature, 3- and 4-adatom nanoclusters are, respectively, revealed in triangular and rectangular patterns, being associated with the large-scale hexagonal symmetry, Bi coverage, and sparsity of vacancies. The stable and metastable structures are consistent with STM measurements of the annealing processes. Furthermore, DOS exhibits a finite value at the Fermi level, a dip structure at $E \sim-0.2$ eV, and peak at $E \sim -0.6$ eV, as identified by the STS measurement. These structures arise from the free conduction electrons, the shifted Dirac-cone, and the bismuth-dominated electronic states, respectively. These features are consistent with those observed by the STS measurements. The stable configuration in room temperature, tunable adatom arrangements, and feature-rich electronic properties may be potentially important for the application of bismuth-adsorbed systems in nanoelectronic devices and energy materials.

%\section{Conclusions}

\par\noindent {\bf Acknowledgments}
This work was supported by the Physics Division, National Center for Theoretical Sciences (South), the National Science Council of Taiwan (Grant Nos. NSC 102-2112-M-006-007-MY3 and 103-2119-M-006-015). We also thank the National Center for High-performance Computing (NCHC) for computer facilities.
%\end{document}

\newpage
%{\large\bf References} \\
\renewcommand{\baselinestretch}{0.2}
%\begin{itemize}
%\newpage

%\end{document}
\newpage \centerline {\Large \textbf {FIGURE CAPTIONS}}

\vskip0.5 truecm %\section{Figure Captures}

Figure 1 Geometric structure of silicon carbide substrate, buffer layer, monolayer rgaphene, and bismuth adatoms. C, Si and Bi are, respectively, indicated by the gray, yellow and purple solid circles.

Figure 2 Ground-state energies of bismuth adsorption on different sites above monolayer graphene.

Figure 3 Geometric structures of hexagonal Bi array by the (a) DFT calculations and (b) STM measurements.

Figure 4 Geometric structures of Bi nanoclusters by the (a) STM measurements and (b) DFT calculations. 3- and 4-adatoms are, respectively, indicated by the green and yellow arrows.

Figure 5 Density of states for hexagonal Bi array by the (a) DFT calculations and (b) STS measurements.

\newpage
\begin{figure}[htb]
\centering\includegraphics[width=14cm]{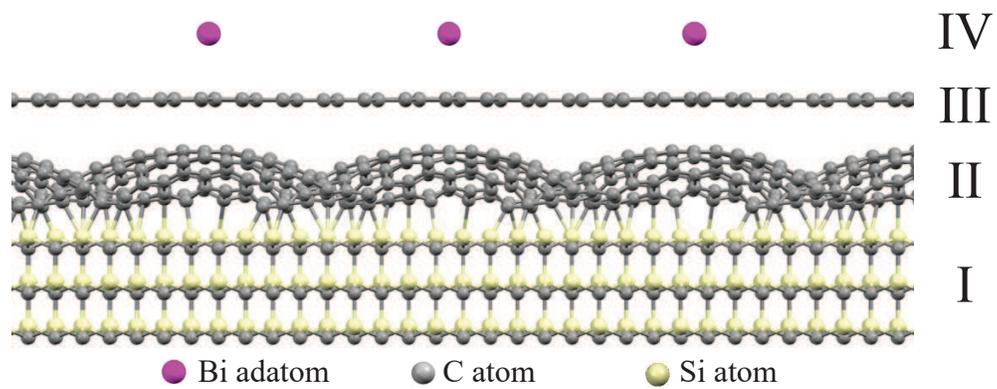}
\caption{Geometric structure of silicon carbide substrate, buffer layer, monolayer rgaphene, and bismuth adatoms. C, Si and Bi are, respectively, indicated by the gray, yellow and purple solid circles.}
\end{figure}

\newpage
\begin{figure}[htb]
\centering\includegraphics[width=14cm]{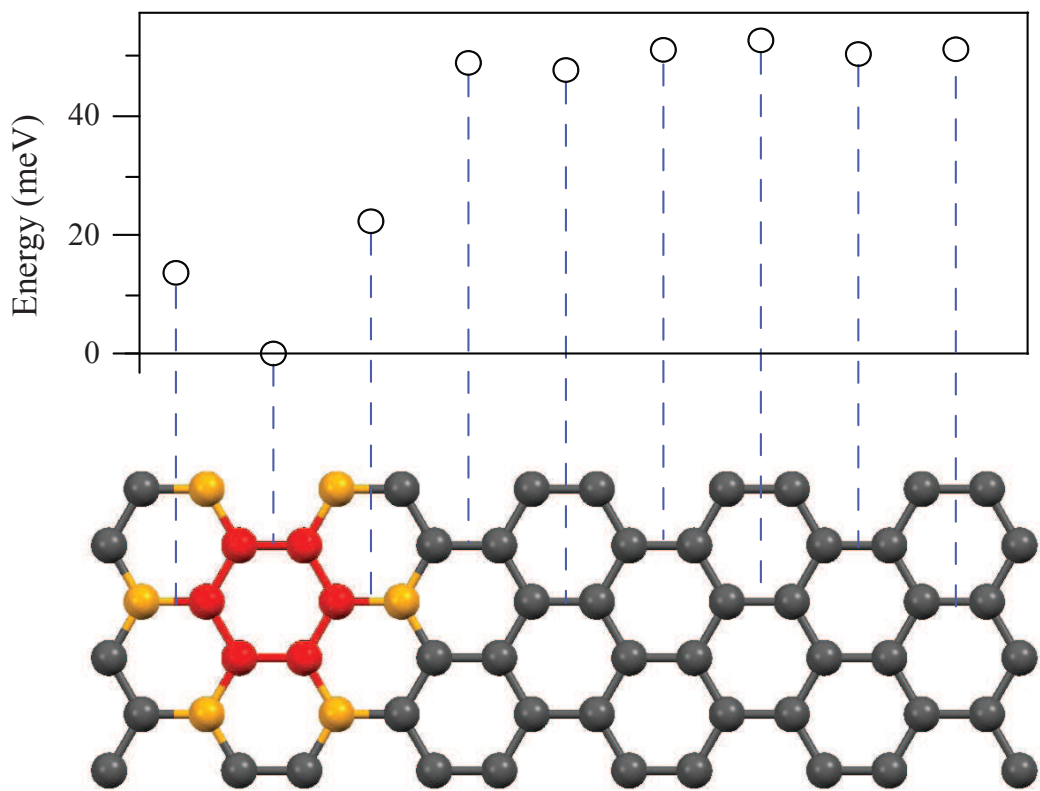}
\caption{Ground-state energies of bismuth adsorption on different sites above monolayer graphene.}
\end{figure}

\newpage
\begin{figure}[htb]
\centering\includegraphics[width=14cm]{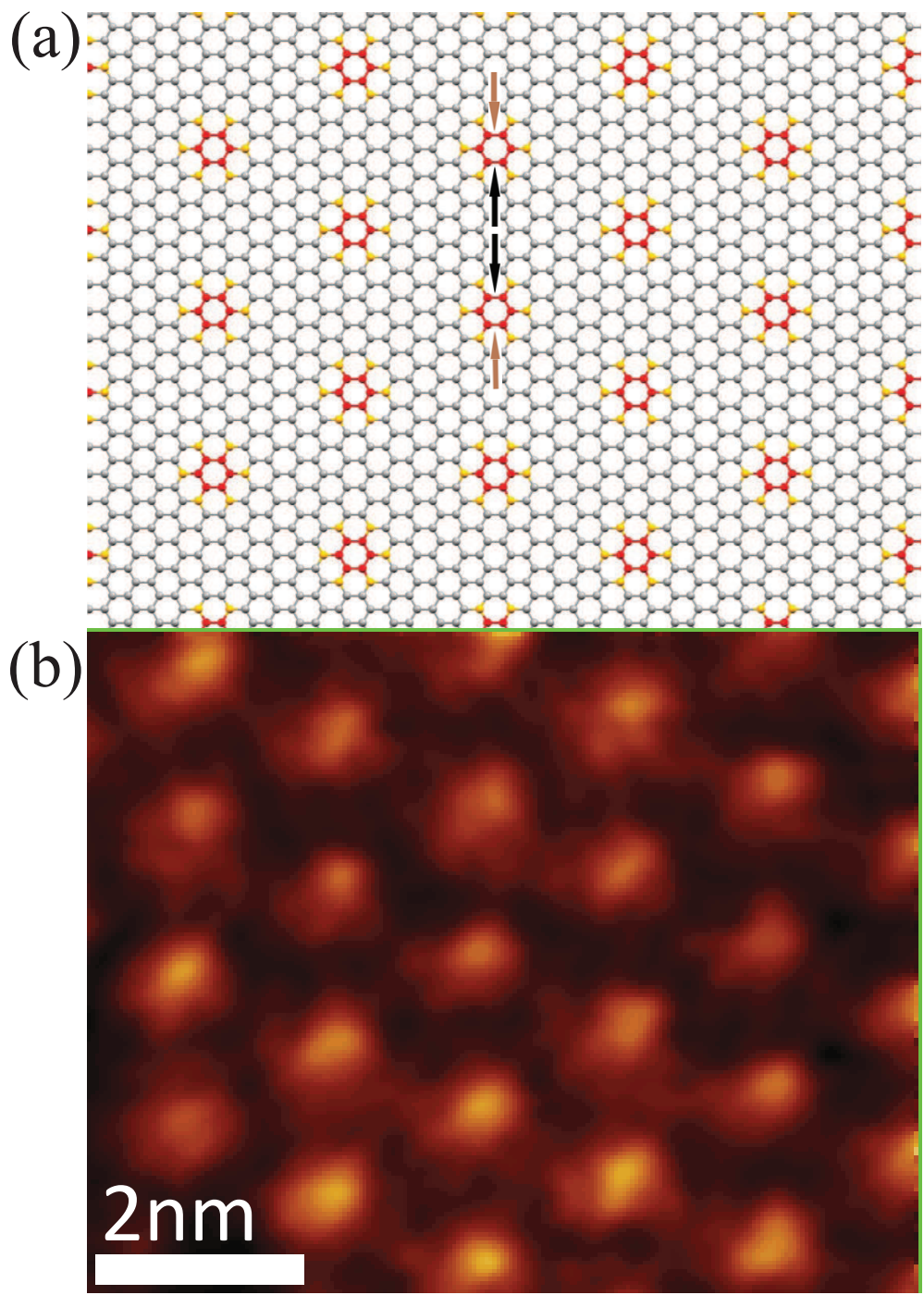}
\caption{Geometric structures of hexagonal Bi array by the (a) DFT calculations and (b) STM measurements.}
\end{figure}

\newpage
\begin{figure}[htb]
\centering\includegraphics[width=14cm]{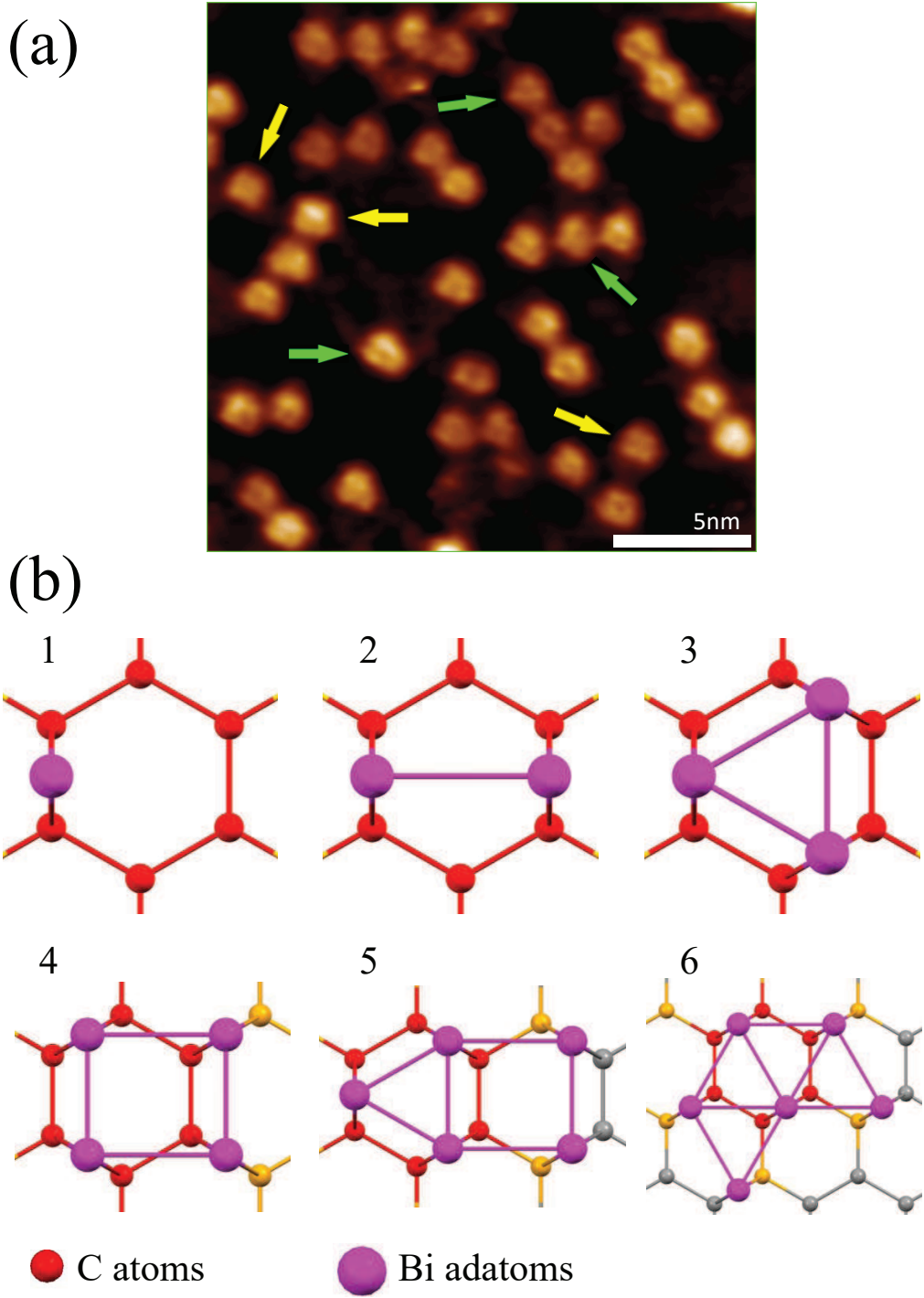}
\caption{Geometric structures of Bi nanoclusters by the (a) STM measurements and (b) DFT calculations. 3- and 4-adatoms are, respectively, indicated by the green and yellow arrows.}
\end{figure}

\newpage
\begin{figure}[htb]
\centering\includegraphics[width=14cm]{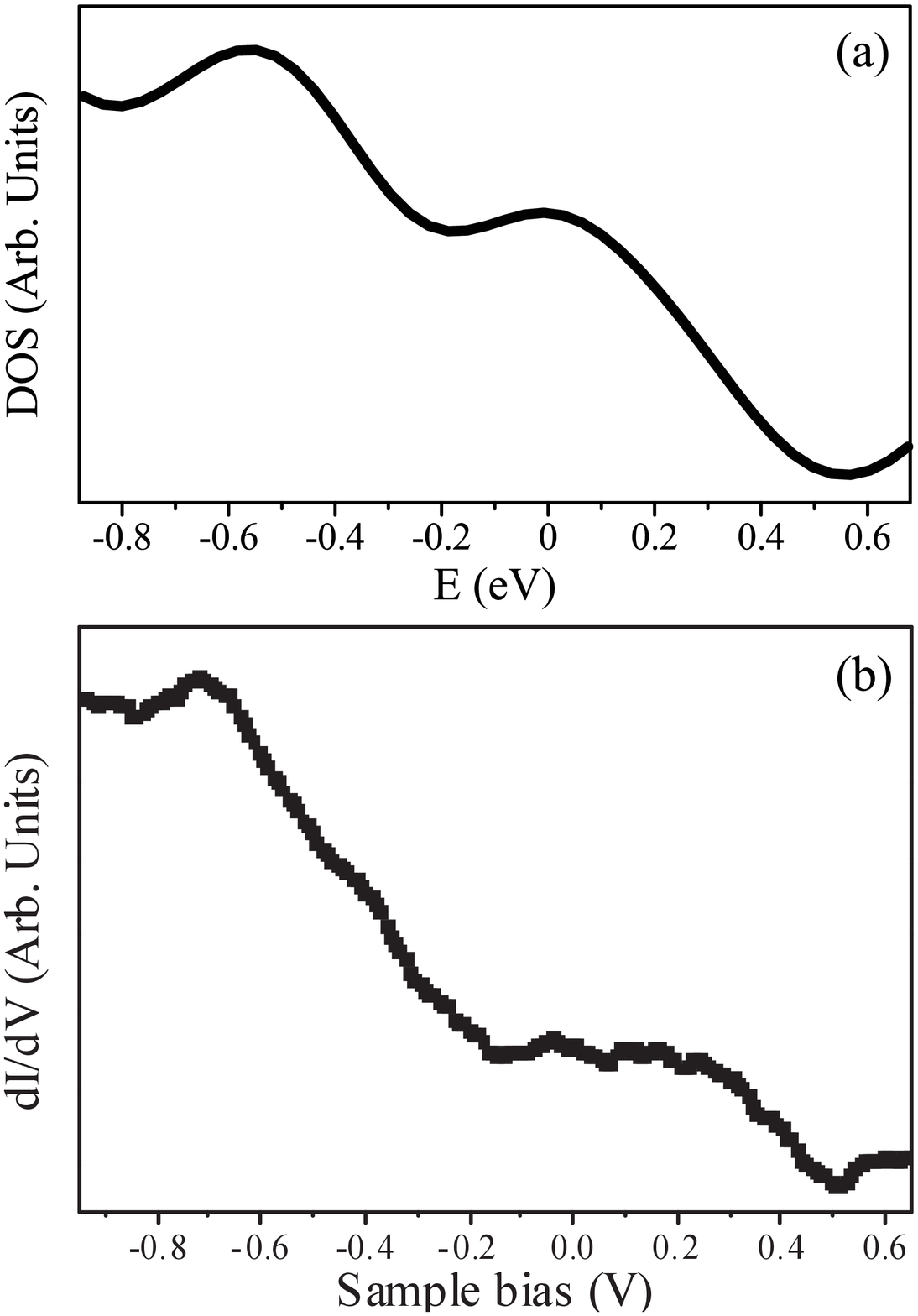}
\caption{Density of states for hexagonal Bi array by the (a) DFT calculations and (b) STS measurements.}
\end{figure}


\begin{thebibliography}{99}
\bibitem{novoselov2004electric}Novoselov, Kostya S., et al. "Electric field effect in atomically thin carbon films." science 306.5696 (2004): 666-669.
\bibitem{geim2007rise}Geim, Andre K., and Konstantin S. Novoselov. "The rise of graphene." Nature materials 6.3 (2007): 183-191.
\bibitem{novoselov2005two}Novoselov, K. S. A., et al. "Two-dimensional gas of massless Dirac fermions in graphene." nature 438.7065 (2005): 197-200.
\bibitem{orlita2008approaching}Orlita, Milan, et al. "Approaching the Dirac point in high-mobility multilayer epitaxial graphene." Physical review letters 101.26 (2008): 267601.
\bibitem{wang2009high}Wang, Shuai, et al. "High mobility, printable, and solution-processed graphene electronics." Nano letters 10.1 (2009): 92-98.
\bibitem{balandin2008superior}Balandin, Alexander A., et al. "Superior thermal conductivity of single-layer graphene." Nano letters 8.3 (2008): 902-907.
\bibitem{barbarino2015intrinsic}Barbarino, Giuliana, Claudio Melis, and Luciano Colombo. "Intrinsic thermal conductivity in monolayer graphene is ultimately upper limited: A direct estimation by atomistic simulations." Physical Review B 91.3 (2015): 035416.
\bibitem{lee2008measurement}Lee, Changgu, et al. "Measurement of the elastic properties and intrinsic strength of monolayer graphene." science 321.5887 (2008): 385-388.
\bibitem{vadukumpully2011flexible}Vadukumpully, Sajini, et al. "Flexible conductive graphene/poly (vinyl chloride) composite thin films with high mechanical strength and thermal stability." Carbon 49.1 (2011): 198-205.
\bibitem{zhang2005experimental}Zhang, Yuanbo, et al. "Experimental observation of the quantum Hall effect and Berry's phase in graphene." Nature 438.7065 (2005): 201-204.
\bibitem{novoselov2006unconventional}Novoselov, K. S., et al. "Unconventional quantum Hall effect and Berry¡¦s phase of 2£k in bilayer graphene." Nature physics 2.3 (2006): 177-180.
\bibitem{min2008room} Min, Hongki, et al. "Room-temperature superfluidity in graphene bilayers." Physical Review B 78.12 (2008): 121401.
\bibitem{perali2013high}Perali, Andrea, David Neilson, and Alex R. Hamilton. "High-temperature superfluidity in double-bilayer graphene." Physical review letters 110.14 (2013): 146803.
\bibitem{neto2009electronic}Neto, AH Castro, et al. "The electronic properties of graphene." Reviews of modern physics 81.1 (2009): 109.
\bibitem{ohta2006controlling}Ohta, Taisuke, et al. "Controlling the electronic structure of bilayer graphene." Science 313.5789 (2006): 951-954.
\bibitem{hwang2007carrier}Hwang, E. H., S. Adam, and S. Das Sarma. "Carrier transport in two-dimensional graphene layers." Physical Review Letters 98.18 (2007): 186806.
\bibitem{sarma2011electronic}Sarma, S. Das, et al. "Electronic transport in two-dimensional graphene." Reviews of Modern Physics 83.2 (2011): 407.
\bibitem{xiao2007valley}Xiao, Di, Wang Yao, and Qian Niu. "Valley-contrasting physics in graphene: magnetic moment and topological transport." Physical Review Letters 99.23 (2007): 236809.
\bibitem{guinea2008midgap}Guinea, F., M. I. Katsnelson, and M. A. H. Vozmediano. "Midgap states and charge inhomogeneities in corrugated graphene." Physical Review B 77.7 (2008): 075422.
\bibitem{lin2015feature}Lin, Shih-Yang, et al. "Feature-rich electronic properties in graphene ripples." Carbon 86 (2015): 207-216.
\bibitem{sutter2009electronic}Sutter, P., et al. "Electronic structure of few-layer epitaxial graphene on Ru (0001)." Nano letters 9.7 (2009): 2654-2660.
\bibitem{hao2010probing}Hao, Yufeng, et al. "Probing Layer Number and Stacking Order of Few-Layer Graphene by Raman Spectroscopy." Small 6.2 (2010): 195-200.
\bibitem{wu2014combined}Wu, Jhao-Ying, Godfrey Gumbs, and Ming-Fa Lin. "Combined effect of stacking and magnetic field on plasmon excitations in bilayer graphene." Physical Review B 89.16 (2014): 165407.
\bibitem{guinea2010energy}Guinea, F., M. I. Katsnelson, and A. K. Geim. "Energy gaps and a zero-field quantum Hall effect in graphene by strain engineering." Nature Physics 6.1 (2010): 30-33.
\bibitem{wong2012strain}Wong, Jen-Hsien, Bi-Ru Wu, and Ming-Fa Lin. "Strain effect on the electronic properties of single layer and bilayer graphene." The Journal of Physical Chemistry C 116.14 (2012): 8271-8277.
\bibitem{castro2007biased}Castro, Eduardo V., et al. "Biased bilayer graphene: semiconductor with a gap tunable by the electric field effect." Physical Review Letters 99.21 (2007): 216802.
\bibitem{lai2008magnetoelectronic}Lai, Y. H., et al. "Magnetoelectronic properties of bilayer Bernal graphene." Physical Review B 77.8 (2008): 085426.
\bibitem{ito2008semiconducting}Ito, Jun, Jun Nakamura, and Akiko Natori. "Semiconducting nature of the oxygen-adsorbed graphene sheet." Journal of applied physics 103.11 (2008): 113712.
\bibitem{ryu2010atmospheric}Ryu, Sunmin, et al. "Atmospheric oxygen binding and hole doping in deformed graphene on a SiO2 substrate." Nano letters 10.12 (2010): 4944-4951.
\bibitem{elias2009control}Elias, D. C., et al. "Control of graphene's properties by reversible hydrogenation: evidence for graphane." Science 323.5914 (2009): 610-613.
\bibitem{ryu2008reversible}Ryu, Sunmin, et al. "Reversible basal plane hydrogenation of graphene." Nano letters 8.12 (2008): 4597-4602.
\bibitem{qu2010nitrogen}Qu, Liangti, et al. "Nitrogen-doped graphene as efficient metal-free electrocatalyst for oxygen reduction in fuel cells." ACS nano 4.3 (2010): 1321-1326.
\bibitem{yeh2014nitrogen}Yeh, Te-Fu, et al. "Nitrogen-Doped Graphene Oxide Quantum Dots as Photocatalysts for Overall Water-Splitting under Visible Light Illumination." Advanced Materials 26.20 (2014): 3297-3303.
\bibitem{krasheninnikov2009embedding}Krasheninnikov, A. V., et al. "Embedding transition-metal atoms in graphene: structure, bonding, and magnetism." Physical review letters 102.12 (2009): 126807.
\bibitem{pi2009electronic}Pi, K., et al. "Electronic doping and scattering by transition metals on graphene." Physical Review B 80.7 (2009): 075406.
\bibitem{machida2006lead}Machida, Motoi, Tomohide Mochimaru, and Hideki Tatsumoto. "Lead (II) adsorption onto the graphene layer of carbonaceous materials in aqueous solution." Carbon 44.13 (2006): 2681-2688.
\bibitem{marchenko2012giant}Marchenko, D., et al. "Giant Rashba splitting in graphene due to hybridization with gold." Nature communications 3 (2012): 1232.
\bibitem{chan2008first}Chan, Kevin T., J. B. Neaton, and Marvin L. Cohen. "First-principles study of metal adatom adsorption on graphene." Physical Review B 77.23 (2008): 235430.
\bibitem{khomyakov2009first}Khomyakov, P. A., et al. "First-principles study of the interaction and charge transfer between graphene and metals." Physical Review B 79.19 (2009): 195425.
\bibitem{balog2010bandgap}Balog, Richard, et al. "Bandgap opening in graphene induced by patterned hydrogen adsorption." Nature materials 9.4 (2010): 315-319.
\bibitem{li2008phase}Li, Lu, et al. "Phase transitions of Dirac electrons in bismuth." Science 321.5888 (2008): 547-550.
\bibitem{hofmann2006surfaces}Hofmann, Ph. "The surfaces of bismuth: Structural and electronic properties." Progress in surface science 81.5 (2006): 191-245.
\bibitem{black2002infrared}Black, M. R., et al. "Infrared absorption in bismuth nanowires resulting from quantum confinement." Physical Review B 65.19 (2002): 195417.
\bibitem{bobaru2012competing}Bobaru, S., et al. "Competing allotropes of Bi deposited on the Al 13 Co 4 (100) alloy surface." Physical Review B 86.21 (2012): 214201.
\bibitem{wanekaya2011applications}Wanekaya, Adam K. "Applications of nanoscale carbon-based materials in heavy metal sensing and detection." Analyst 136.21 (2011): 4383-4391.
\bibitem{shan2009polycrystalline}Shan, Dan, et al. "Polycrystalline bismuth oxide films for development of amperometric biosensor for phenolic compounds." Biosensors and Bioelectronics 24.12 (2009): 3671-3676.
\bibitem{li2013bismuth}Li, Yuling, et al. "Bismuth oxide: a new lithium-ion battery anode." Journal of Materials Chemistry A 1.39 (2013): 12123-12127.
\bibitem{akturk2010bismuth}Akturk, Olcay Uzengi, and Mehmet Tomak. "Bismuth doping of graphene." Applied Physics Letters 96.8 (2010): 081914.
\bibitem{hsu2013first}Hsu, Chia-Hsiu, Vidvuds Ozolins, and Feng-Chuan Chuang. "First-principles study of Bi and Sb intercalated graphene on SiC (0001) substrate." Surface Science 616 (2013): 149-154.
\bibitem{chen2015long}Chen, H-H., et al. "Long-range interactions of bismuth growth on monolayer epitaxial graphene at room temperature." Carbon 93 (2015): 180-186.
\bibitem{chen2015tailoring}Chen, H-H., et al. "Tailoring low-dimensional structures of bismuth on monolayer epitaxial graphene." Scientific reports 5 (2015).
\bibitem{kresse1996efficient} Kresse, Georg, and Jurgen Furthmuller. "Efficient iterative schemes for ab initio total-energy calculations using a plane-wave basis set." Physical Review B 54.16 (1996): 11169.
\bibitem{perdew1996generalized}Perdew, John P., Kieron Burke, and Matthias Ernzerhof. "Generalized gradient approximation made simple." Physical review letters 77.18 (1996): 3865.
\bibitem{grimme2006semiempirical}Grimme, Stefan. "Semiempirical GGA-type density functional constructed with a long?range dispersion correction." Journal of computational chemistry 27.15 (2006): 1787-1799.
\end{thebibliography}
\end{document}